\documentclass[12pt]{iopart}
\usepackage{iopams}
\usepackage{cite}
\usepackage{graphicx}
\usepackage{titlesec}
\usepackage{fancyhdr}
\usepackage{indentfirst}
\begin{document}

\title{Nuclear chart in covariant density functional theory with dynamical correlations: From Oxygen to Tin}
\fancyhf{}
\pagestyle{fancy}
\author{Yi-Long Yang and Ya-Kun Wang}

\address{State Key Laboratory of Nuclear Physics and Technology, School of Physics, Peking University, Beijing 100871, China}
\ead{wangyk15@pku.edu.cn}
\vspace{10pt}
\begin{indented}
\item[]September 2019
\end{indented}

\begin{abstract}
    Nuclear masses of even-even nuclei with the proton number $8\leq Z\leq 50$ (O to Sn isotopes) from proton drip line to neutron drip line are investigated using the triaxial relativistic Hartree-Bogoliubov (RHB) theory with the relativistic density functional PC-PK1, and the dynamical correlation energies (DCEs) associated with the rotational motion and the quadrupole shape vibrational motion are taken into account by the five-dimensional collective Hamiltonian (5DCH) method.
    The root-mean-square deviation with respect to the experimental masses reduces from 2.50 MeV to 1.59 MeV after the consideration of DCEs.
    The inclusion of DCEs has little influence on the position of drip lines, and the predicted numbers of bound even-even nuclei between proton and neutron drip lines from O to Sn isotopes are respectively 569 and 564 for the cases with and without DCEs.
\end{abstract}
\section{Introduction}
Nuclear mass is one of the most fundamental properties of nuclei.
It is of great importance not only in the nuclear physics but also in the astrophysics~\cite{Lunney2003RMP, Blaum2006PR}.
For example, the masses of nuclei widely ranged from the valley of stability to the vicinity of neutron drip line are involved in simulating the rapid neutron capture (r-process) of stellar nucleosynthesis~\cite{Arnould2007PR}.
Although considerable achievements have been made in mass measurement~\cite{Wang2017CPC}, vast amount of nuclei on the neutron-rich side away from valley of stability are still beyond the experimental capability in the foreseeable future.
Therefore, reliable nuclear models for high-precision description of nuclear masses are strongly required.

During the past decades, various global nuclear models have been proposed to describe the nuclear mass, including the finite-range droplet model (FRDM)~\cite{Moller1995ADaNDT, Moeller2016ADNDT}, the semi-empirical Weizs$\mathrm{\ddot{a}}$cker-Skyrme (WS) model~\cite{Wang2010PRC, Wang2014PLB}, the non-relativistic~\cite{Goriely2009PRLa, Goriely2013PRC, Erler2012N, Goriely2016TheEuropeanPhysicalJournalA52202} and relativistic~\cite{Geng2005PoTP, Afanasjev2013PLB, Agbemava2014PRC,Zhang2014FoP, Lu2015PRC} density functional theories (DFTs), etc.
The nuclear DFTs start from universal density functionals containing a few parameters determined by fitting to the properties of finite nuclei or nuclear matter.
They can describe the nuclear masses, ground and excited state properties in a unified way~\cite{Ring1996PPNP, Bender2003RMP, Vretenar2005PR}.
In particular, due to the consideration of the Lorentz symmetry, the relativistic or covariant density functional theory (CDFT) naturally includes the nucleonic spin degree of freedom and the time-odd mean fields, which play an essential role in describing the moments of inertia for nuclear rotations~\cite{Koenig1993PRL, Afanasjev2000NPA, Afanasjev2000PRC, Zhao2012PRC}.
Up to now, the CDFT has received wide attentions because of its successful description of many nuclear phenomena~\cite{Ring1996PPNP, Vretenar2005PR, Meng2006PiPaNP, Niksic2011PiPaNP, Meng2013FoP, meng2016relativistic, Zhao2018Int.J.Mod.Phys.E1830007}.

In the framework of CDFT, the masses for over 7000 nuclei with $8\leq Z\leq100$ up to the proton and neutron drip lines were investigated based on the axial relativistic mean field (RMF) theory~\cite{Geng2005PoTP}.
Later on, to explore the location of the proton and neutron drip lines, a systematic investigation has been performed for even-even nuclei within the axial relativistic Hartree-Bogoliubov (RHB) theory~\cite{Afanasjev2013PLB, Agbemava2014PRC, Afanasjev2015PRC}.
Very recently, the ground-state properties of nuclei with $8\leq Z\leq 120$ from the proton drip line to the neutron drip line have been calculated using the spherical relativistic continuum Hartree-Bogoliubov (RCHB) theory, in which the couplings between the bound states and the continuum can be considered properly~\cite{Xia2018ADaNDT}.
The root-mean-square (rms) deviation with respect to the experimental nuclear masses in these pure CDFT calculation is typically around several MeV.
To achieve a higher precision, one needs to go beyond the mean-field approximation and consider the beyond-mean-field dynamical correlation energies (DCEs).

In Ref.~\cite{Zhang2014FoP}, Zhang \emph{et al.} have carried out a global calculation of the binding energies for 575 even-even nuclei ranging from $Z=8$ to $Z=108$ based on the axial RMF, and the Bardeen-Cooper-Schrieffer (BCS) approximation is adopted to consider the pairing correlations.
In this axial RMF+BCS calculation, the DCEs, namely the rotational and vibrational correlation energies were obtained by cranking prescription.
After including the DCEs, the rms deviation for binding energies of the 575 even-even nulcei reduces from 2.58 MeV to 1.24 MeV.
Later on, the DCEs of these nuclei were revisited in Ref.~\cite{Lu2015PRC} using the five-dimensional collective Hamiltonian (5DCH) method with the collective parameters determined by the CDFT calculations~\cite{Niksic2009PRC, Li2009PRC}.
The 5DCH method takes into account the DCEs in a more proper way, and the resulting rms deviation reduces from 2.52 MeV to 1.14 MeV~\footnote{Note that in Ref.~\cite{Zhang2014FoP}, the adopted experimental mass data are taken from Ref.~\cite{Audi2003NuclearPhysicsA729337-676}, while in Ref.~\cite{Lu2015PRC}, the experimental mass data are from Ref.~\cite{Audi2012ChinesePhysicsC361287--1602}.
Moreover, compared to the theoretical results shown in Ref.~\cite{Zhang2014FoP}, the energies associated with triaxial deformation are further included in Ref.~\cite{Lu2015PRC}.}

The studies shown in Refs.~\cite{Zhang2014FoP, Lu2015PRC} demonstrate that the inclusion of the DCEs can significantly improve the description of the nuclear masses.
So far, the inclusion of DCEs in the CDFT is still confined to the nuclei with mass known and the DCEs of most neutron-rich nuclei crucial in simulating the r-process still remain uninvestigated.
Therefore, it is necessary to extend the investigation from the nuclei with mass known to the boundary of nuclear landscape.
Meanwhile, the pairing correlations are treated by BCS approximation in Refs.~\cite{Zhang2014FoP, Lu2015PRC}.
For the description of nuclei around the neutron drip line, this approximation is questionable because the continuum effect can not be taken into account properly ~\cite{Dobaczewski1984NPA}.
Nevertheless, the methods with the Bogoliubov transformation can provide a better description for the pairing correlations in weakly bound nuclei.
Therefore, in present paper, nuclear masses of even-even nuclei from O to Sn isotopes ranging from the proton drip line to the neutron drip line are performed within the triaxial relativistic Hartree-Bogoliubov theory~\cite{PhysRevC.81.054318}, and the beyond mean-field quadrupole DCEs are included by the 5DCH method.

\section{Numerical details}
To this end, the large-scale deformation constrained triaxial RHB calculation is carried out to generate the mean-filed states in the whole $(\beta,\gamma)$ plane.
The well-known density functional PC-PK1~\cite{Zhao2010PRC} is adopted in the particle-hole channel.
This density functional particularly improves the description for isospin dependence of binding energies and it has been successfully used in describing the Coulomb displacement energies between mirror nuclei~\cite{Sun2011Sci.ChinaSer.G-Phys.Mech.Astron.210}, nuclear masses~\cite{Zhao2012Phys.Rev.C64324,Lu2015PRC}, quadrupole moments~\cite{Zhao2014Phys.Rev.C11301,Yordanov2016Phys.Rev.Lett.32501,Haas2017EPL62001}, superheavy nuclei~\cite{Zhang2013Phys.Rev.C54324,Lu2014Phys.Rev.C14323,Agbemava2015Phys.Rev.C54310,Li2015Front.Phys.268}, nuclear shape phase transitions~\cite{Quan2018Phys.Rev.C31301, Quan2017PRC}, magnetic and antimagnetic rotations~\cite{Zhao2011PRL,Zhao2011PLB,Meng2013FoP,Meng2016PhysicaScripta53008}, chiral rotations~\cite{Zhao2017Phys.Lett.B1}, etc.
A finite range separable pairing force with $G=-728$ MeV~\cite{Tian2009PLB} is used in the particle-particle channel.
The triaxl RHB equation is solved by a three-dimensional harmonic oscillator basis expansion in Cartesian coordinate with 12 and 14 major shells for nuclei with $Z<20$ and $20\leq Z\leq50$, respectively.
The obtained quasiparticle energies and wave functions are used to calculated the mass parameters, moments of inertia, and collective potentials in the 5DCH, which are functions of the quadrupole deformation parameters $\beta$ and $\gamma$.
The DCE $E_{\mathrm{corr}}$ is defined as the energy difference between the lowest mean-field state and the $0_1^+$ state of 5DCH.

\section{Results and discussion}
\begin{figure}[h]
  \centering
  \includegraphics[width=11cm]{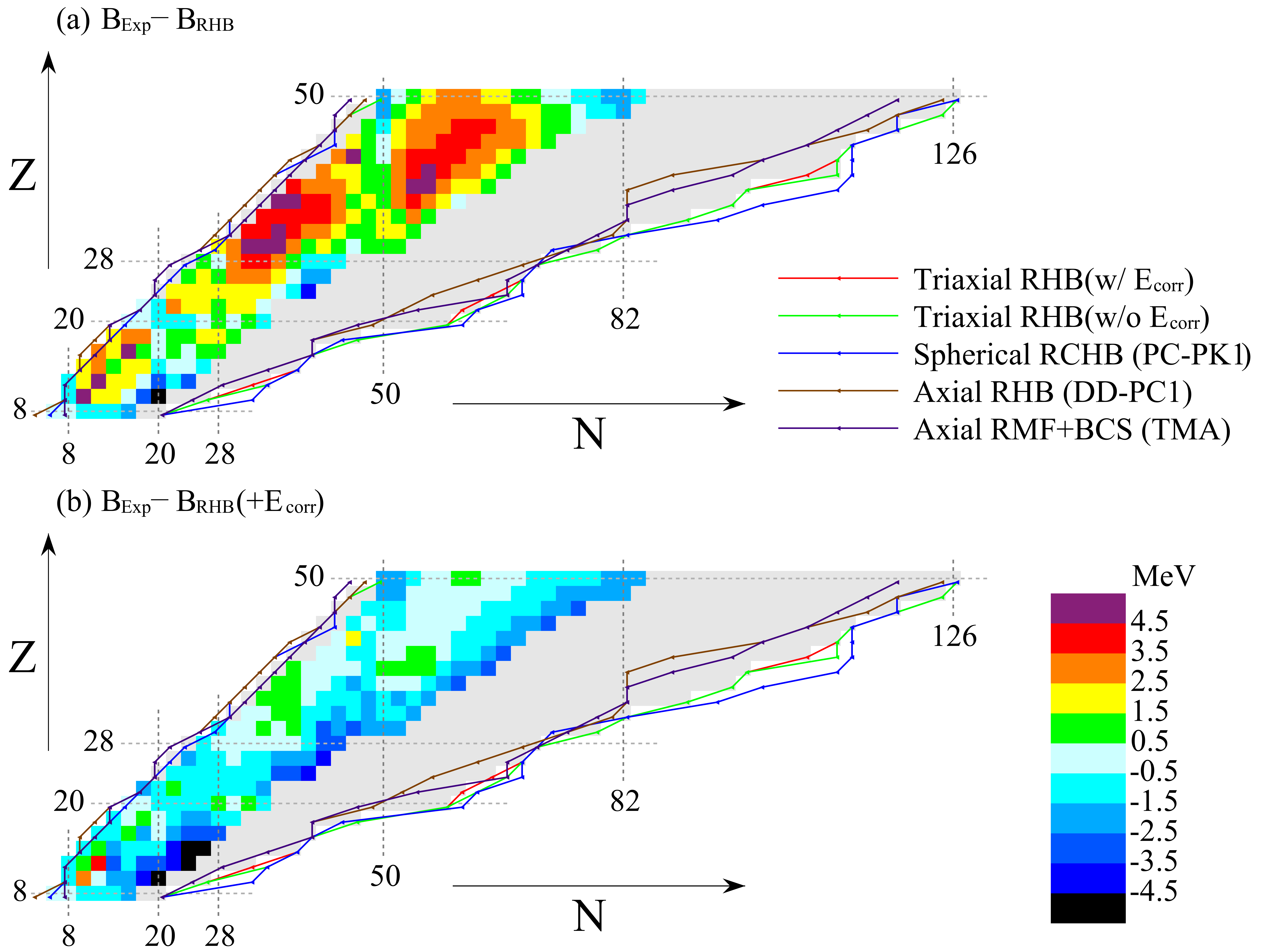}\\
  \caption{Even-even nuclei from O to Sn isotopes predicted by the triaxial RHB approach with (panel (b)) and without (panel (a)) dynamical correlation energies.
  Discrepancies of the calculated binding energies with the data~\cite{Wang2017CPC} are denoted by colors.
  The proton and neutron drip lines predicted by spherical RCHB (PC-PK1)~\cite{Xia2018ADaNDT}, axial RHB (DD-PC1)~\cite{Agbemava2014PRC} and axial RMF+BCS (TMA)~\cite{Geng2005PoTP} are also plotted for comparison.}\label{1}
\end{figure}
The bound nuclear regions from O to Sn isotopes predicted by the triaxial RHB approach with and without DCEs are shown in Figure~\ref{1}.
The discrepancies of the calculated binding energies with respect to the data are scaled by colors.
The binding energies calculated by triaxial RHB approach shown in panel (a) are given by the binding energies of the lowest mean-field states, while in panel (b), the DCEs are taken into account.

In the triaxial RHB calculations without DCEs, it is found that the binding energies are systematically underestimated.
Most of the deviations are in the range of 0.5 MeV $\sim$ 4.5 MeV, resulting in the rms deviation of 2.50 MeV.
By including the DCEs, the underestimation of binding energies are improved significantly, and the rms deviation is reduced from 2.50 MeV to 1.59 MeV.
However, in the region $(N,Z)\sim(24,12)$, large deviations exist even though the DCEs have been considered.
This might be associated with the complex shell evolution around this region.
To have a better description of binding energies in this region, the tensor interaction~\cite{Long2007Phys.Rev.C34314} may need to be included in the adopted density functional, which is beyond the scope of the present investigations.

In order to estimate the number of bound nuclei from O to Sn isotopes, two-proton and two-neutron drip lines predicted by present triaxial RHB approach with and without DCEs are also plotted in Figure.~\ref{1}.
The predicted number of bound even-even nuclei between proton and neutron drip lines from O to Sn isotopes without DCEs is 569.
The inclusion of DCEs has little impact on the proton and neutron drip lines and the corresponding number of bound nuclei is 564.
For comparison, the drip lines predicted by the spherical RCHB (PC-PK1)~\cite{Xia2018ADaNDT}, axial RHB(DD-PC1)~\cite{Agbemava2014PRC} and axial RMF+BCS(TMA)~\cite{Geng2005PoTP} are also shown.
It is found that theoretical differences for proton drip lines are rather small.
However, the neutron drip lines predicted by different approaches differ considerably and the differences increase with the mass number.
The neutron drip line predicted by the triaxial RHB approach locates in between of those by the axial RHB and spherical RCHB approaches.

\begin{figure}[h]
  \centering
  \includegraphics[width=11cm]{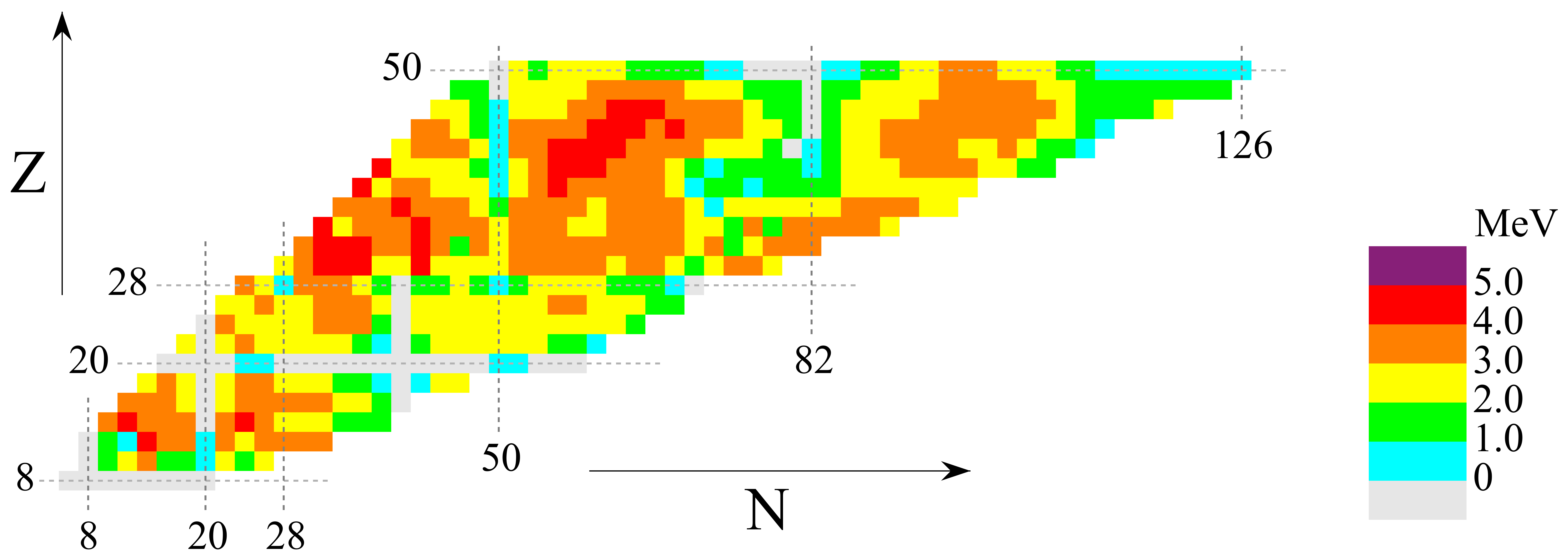}\\
  \caption{Contour map of the dynamical correlation energies $E_\mathrm{corr}$ calculated by the 5DCH model based on triaxial RHB calculations as functions of neutron and proton numbers.}\label{2}
\end{figure}
Figure \ref{2} displays the contour map of the dynamical correlation energies $E_\mathrm{corr}$ calculated by the 5DCH model based on triaxial RHB calculations.
The calculated $E_\mathrm{corr}$ ranges from 0 to 5 MeV, and varies mainly in the region of $2.0$---$4.0$ MeV.
Due to the shape fluctuations, the dynamical correlation energies are pronounced for nuclei around $Z \sim 32, 40$ and $N \sim 34, 60$.
Similarly to the finding reported in Ref.~\cite{Lu2015PRC}, the dynamical correlation energies for the semi-magic nuclei with $Z=28, 50$ and $N = 28, 82$ are nonzero or even rather large.
This is caused by the fact that the potential energy surfaces for these nuclei are either soft or with shape coexisting phenomena.

\begin{figure}[h]
  \centering
  \includegraphics[width=10cm]{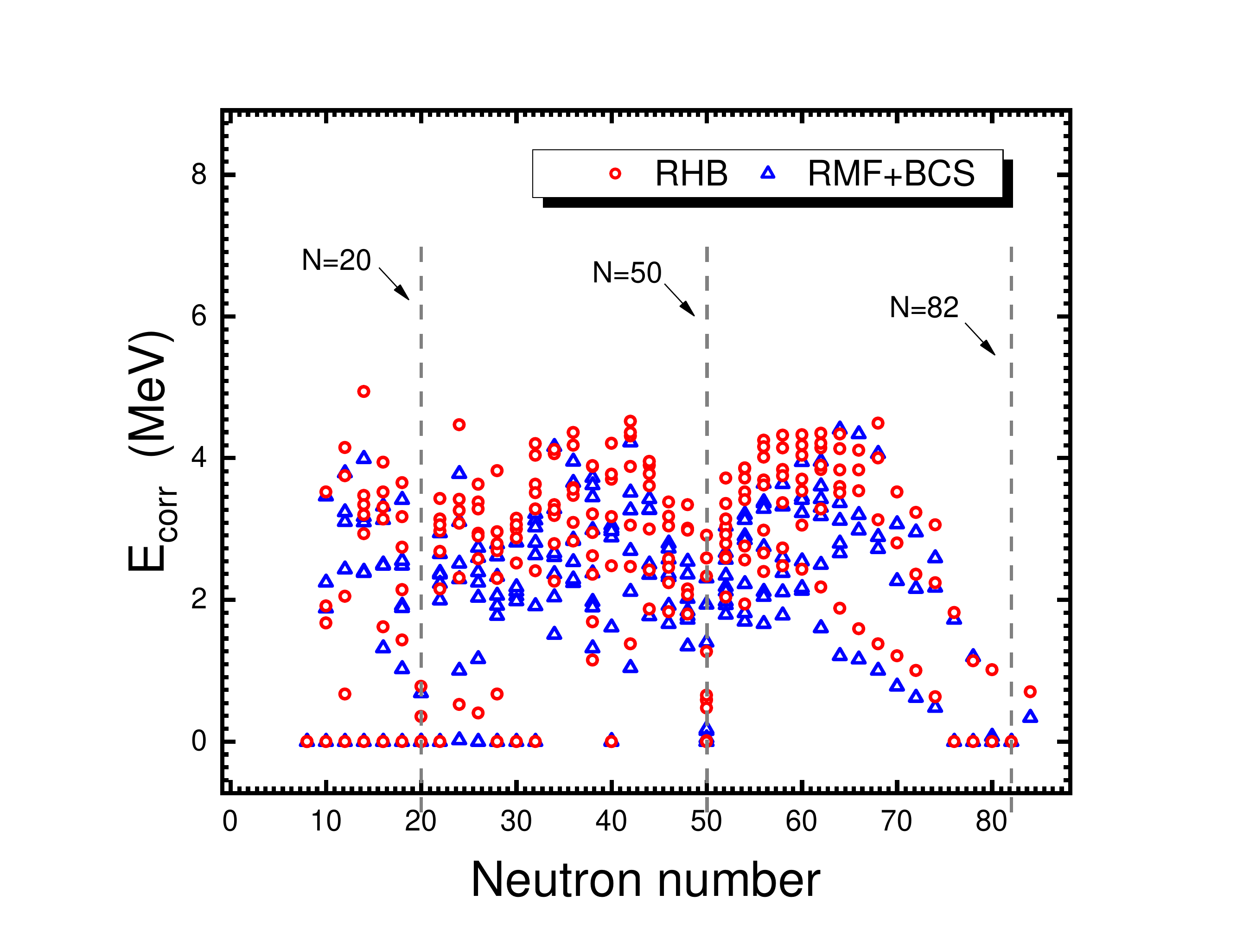}\\
  \caption{The dynamical correlation energies calculated by 5DCH based on triaxial RHB (circles) in comparison with those based on triaxial RMF+BCS~\cite{Lu2015PRC}  (triangles).}\label{3}
\end{figure}
In Ref.~\cite{Lu2015PRC}, the binding energies of 575 even-even nuclei in the region of $8\leq Z\leq108$ have been calculated using the 5DCH method in the framework of triaxial RMF+BCS.
For the 228 nuclei with $8\leq Z\leq50$ in Ref.~\cite{Lu2015PRC}, the rms deviation with respect to data is 1.23 MeV, whereas the rms deviation in present calculations for these nuclei is 1.47 MeV.
It is found that the lowest mean-field binding energies given by these two calculations are of little difference, so the differences mainly come from the $E_{\mathrm{corr}}$.

The dynamical correlation energies $E_{\mathrm{corr}}$ calculated by 5DCH based on triaxial RHB and triaxial RMF+BCS are plotted in Figure \ref{3} as functions of neutron number $N$.
Even though the systematics of $E_{\mathrm{corr}}$ are similar for both calculations, the triaxial RHB-based $E_{\mathrm{corr}}$ are systematically larger than those based on triaxial RMF+BCS.
The rms deviation between these two results is 0.53 MeV and this leads to the overall difference in the binding energies.
The systematic difference of $E_{\mathrm{corr}}$ might be originated from the different treatments of pairing correlations.
The pairing correlations in present calculations are considered by the Bogoliubov transformation, while in Ref.~\cite{Lu2015PRC}, the pairing correlations are considered by the BCS approximation.
Different pairing properties may lead to different zero point energies, and thus results in different dynamical correlation energies $E_{\mathrm{corr}}$~\cite{Lu2015PRC}.

\begin{figure}[h]
  \centering
  \includegraphics[width=11cm]{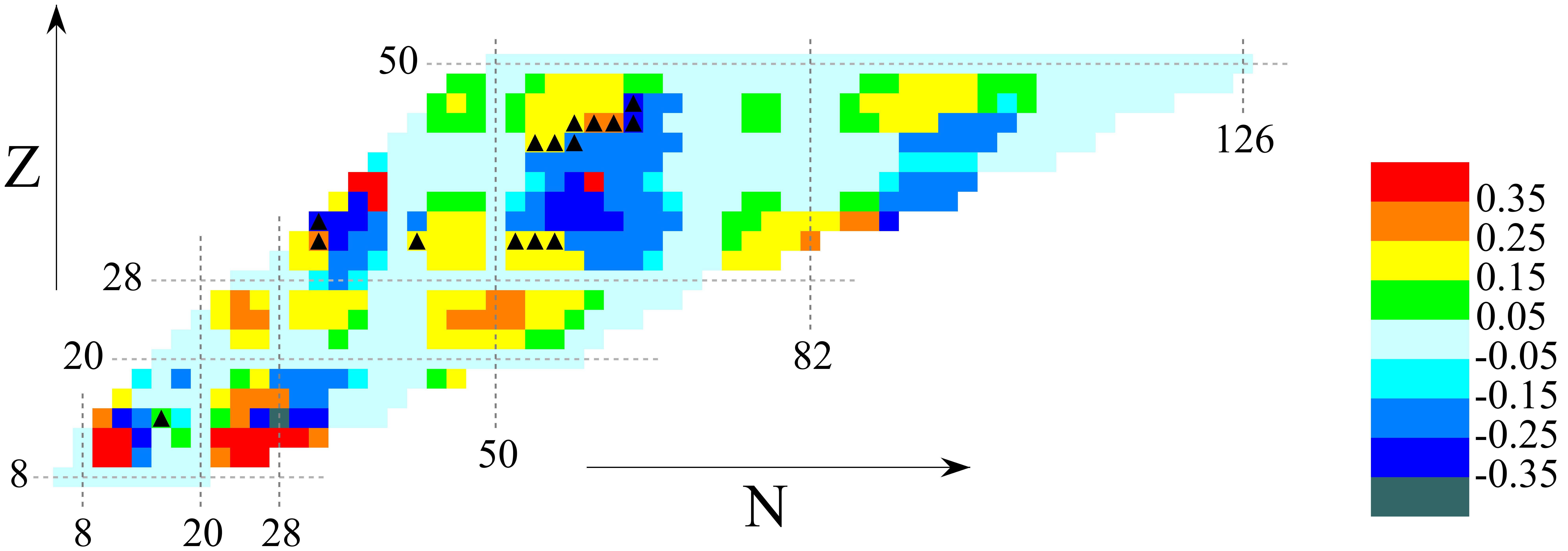}\\
  \caption{Contour map of the quadrupole deformation $\beta$ calculated by the triaxial RHB approach as functions of the neutron and proton numbers. Nuclei with triaxial deformation are denoted by black triangles.}
  \label{4}
\end{figure}
The contour map of triaxial RHB calculated quadrupole deformation $\beta$ are presented in Figure \ref{4}.
The quadrupole deformation corresponds to the energy minima on the whole $(\beta,\gamma)$ plane.
Here, $\beta$ is defined as positive for $0^\circ\leq\gamma<30^\circ$ and negative for $30^\circ<\gamma\leq60^\circ$.
In general, the nuclei near magic numbers possess small or vanishing deformation.
However, it is found that single magic numbers do not enforce sphericity, especially for neutron-rich nuclei.
For example, neutron-rich isotones with $N = 28, Z<20$ and $N = 50, Z<28$ show remarkable deformation.
In addition, the deformation develops when moving away from the magic numbers either isotopically or isotonically.
There are four large deformed regions located at $(N, Z)\sim (24,14), (34,32), (60,40)$ and $(94,46)$.
These regions with large deformation correspond to the regions with large DCEs as shown in Figure~\ref{2}.

The nuclei with triaxial deformation i.e. $\gamma\neq 0^\circ, 60^\circ$, are also shown in Figure~\ref{4}.
There are 15 nuclei with triaxial deformation and most of them belong to Ge, Mo, and Ru isotopes.
Our theoretical investigations provide good candidates for the experiment to study the possibility of triaxial deformations.

\section{Summary}
In summary, the nuclear masses of even-even nuclei with $8\leq Z\leq 50$ ranging from the proton drip line to neutron drip line are systematically investigated using the triaxial relativistic Hartree-Bogoliubov theory with the relativistic density functional PC-PK1, and the quadrupole dynamical correlation energies are taken into account by solving the five-dimensional collective Hamiltonian.
With the inclusion of dynamical correlation energies, the prediction of triaxial relativistic Hartree-Bogoliubov theory for 252 nuclei masses is improved significantly with the rms deviation reducing from 2.50 MeV to 1.59 MeV.
It is found that the dynamical correlation energies have little influence on the positions of proton and neutron drip lines, and the predicted numbers of bound even-even nuclei between proton and neutron drip lines with and without dynamical correlation energies are 569 and 564 respectively.
In present calculations, the obtained dynamical correlation energies range from 0 to 5 MeV which are slightly larger than the results of previous work~\cite{Lu2015PRC}.
The discrepancies might be caused by the different treatments of pairing correlations, which would lead to different zero point energies, and thus different dynamical correlation energies.
The contour map of quadrupole deformation $\beta$ and $\gamma$ associated with the dynamical correlation energies is also discussed in detail.
There are 15 nuclei predicted to have triaxial deformation, which provide good candidates for the experiment to study the possibility of triaxial deformations.
The final aim of this project is to build a whole nuclear mass table including both triaxial degrees of freedom and dynamical correlation energies. The works following this line are still in progress.

\ack
The authors are grateful to Prof. Zhi Pan Li and Prof. PengWei Zhao for providing
the numerical computation codes and the fruitful discussions as well as critical readings
of our manuscript.
This work was partly supported by the National Key R\&D Program of China (Contract No. 2018YFA0404400 and No. 2017YFE0116700) and
the National Natural Science Foundation of China (Grants No. 11621131001, No. 11875075, No. 11935003, and No. 11975031).

\end{document}